\documentclass[12pt]{article}
\usepackage{hyperref}
\usepackage[english]{babel}
\usepackage{amsmath}
\usepackage{latexsym}
\usepackage{amssymb}
\usepackage[all]{xy}
\usepackage[dvips]{graphicx}
\usepackage{epsfig}
\usepackage{psfrag}

\numberwithin{equation}{section} \numberwithin{table}{section}
\numberwithin{figure}{section}

\begin{document}



\begin{titlepage}
   \begin{flushright}
{\small MPP-2013-296 }
  \end{flushright}

   \begin{center}

     \vspace{20mm}

     {\LARGE \bf Holographic charge transport in 2+1 \vspace{1mm}dimensions at finite $N$
     \vspace{3mm}}

     \vspace{10mm}

    Shan Bai${}^{a}$, Da-Wei Pang${}^{b}$\\
     \vspace{5mm}
      {\small \sl ${}^{a}$Theoretisch-Physikalisches Institut, Friedrich-Schiller-Universit\"{a}t Jena,\\
             Max-Wien-Platz 1, 07743 Jena, Germany\\
      \sl ${}^{b}$Max-Planck-Institut f\"{u}r Physik (Werner-Heisenberg-Institut)\\
      F\"{o}hringer Ring 6, 80805 M\"{u}nchen, Germany}\\

     {\small \tt shan.bai@uni-jena.de, dwpang@mppmu.mpg.de}
     \vspace{10mm}

   \end{center}

\begin{abstract}
\baselineskip=18pt
We study holographic charge transport in (2+1) dimensions at finite $N$, whose dual gravity background is given by perturbative black hole solution in Einstein theory plus cubic terms of Weyl tensor. We consider the higher derivative corrections to the standard Maxwell action, given by the interacting terms between the Weyl tensor and the field strength. We calculate the DC conductivity by using both the membrane paradigm and the Kubo's formula and find precise agreement. We compute the AC conductivity and find an analog of the crossover from `metal' to `bad metal' in the low frequency limit. Moreover, the conductivity becomes a constant in the large frequency limit. We derive two universal relations for the Green's functions and observe that they are exactly the same as the infinite $N$ counterparts.
\end{abstract}
\setcounter{page}{0}
\end{titlepage}

\pagestyle{plain} \baselineskip=19pt

\tableofcontents

\section{Introduction}
The AdS/CFT correspondence has been proven to be a powerful tool for studying the dynamics of strongly coupled quantum field theories~\cite{Maldacena:1997re, GKP:1998, Witten:1998, Aharony:1999ti}. Recently there have been considerable successful applications of AdS/CFT to condensed matter theories (AdS/CMT). For reviews on this topic, see~\cite{Hartnoll:2009sz, Herzog:2009, McGreevy:2009, Horowitz:2010, Sachdev:2010, Hartnoll:2011}. The AdS/CFT correspondence in the usual sense amounts to taking infinite $N$ and infinite 't Hooft coupling, while the dual gravity theory is given by Einstein gravity. Furthermore, adding higher derivative terms in the bulk gravity side allows us to probe the finite $N$ effects in the dual field theory side. One novel example of such higher derivative gravity theories is Gauss-Bonnet gravity, where exact black hole solutions are explicitly known~\cite{Cai:2001dz}. It has been widely recognized that higher derivative gravity can change the physics qualitatively, for instance, addition of Gauss-Bonnet terms makes the transition to the superconducting phase more difficult~\cite{Gregory:2009fj}.

AdS/CMT indicates that the dual gravity lives in $(3+1)$ dimensional bulk spacetime, in which the Gauss-Bonnet terms become trivial. Therefore in order to study the finite $N$ effects in AdS/CMT, one has to consider higher derivative corrections to 4D gravity theories other than the Gauss-Bonnet terms. Recently higher derivative effects in 4D AdS spacetimes were extensively investigated in~\cite{Smolic:2013gz}, where perturbative black hole solutions in various 4D higher derivative gravity theories were obtained. These black hole solutions provide interesting backgrounds for studying finite $N$ effects in AdS/CMT.

The main focus of this paper is to study holographic charge transport in $(2+1)$ dimensions at finite $N$, whose dual gravity background is the perturbative black hole solution in Einstein gravity plus higher derivative corrections obtained in~\cite{Smolic:2013gz}. Charge transport in strongly-interacting quantum critical systems, where well-defined quasiparticle excitations are absent, is a very interesting problem in condensed matter physics. For systems with well-defined quasiparticle excitations, the collective dynamics is effectively determined by a quantum Boltzmann equation. Many equilibrium properties, such as the electrical conductivity and thermal transport, can be described by the Boltzmann equation. One notable example is the $O(N)$ model in the large $N$ limit~\cite{Sachdev:2011qpt, Sachdev:1997prb}, where the conductivity exhibits a Drude peak in the low frequency limit and asymptotes to a universal constant $\sigma_{\infty}$ at high frequencies $\omega\gg T$. Note that the limits $\omega\rightarrow0$ and $T\rightarrow0$ do not commute when discussing the low frequency, low temperature properties of the conductivity, which plays a key role in extracting the correct behavior~\cite{Damle:1997prb}.

Since the AdS/CFT correspondence is a strong-weak duality, it provides a useful framework of characterizing the properties of quantum critical systems. The charge transport near quantum critical points in AdS/CFT was initiated in~\cite{Herzog:2007ij}, where the dual gravity description was a stack of black M2-branes. It was shown in~\cite{Herzog:2007ij} that the following relation holds for generic temperature $T$,
\begin{equation}
\label{KLKT}
K^{L}(\omega,k)K^{T}(\omega,k)={\rm const},
\end{equation}
where $K^{L}(\omega,k)$ and $K^{T}(\omega,k)$ are related to the components of the current-current correlation functions as follows
\begin{equation}
G^{R}_{tt}(\omega,k)=-\frac{k^{2}}{\sqrt{k^{2}-\omega^{2}}}K^{L}(\omega,k),~~
G^{R}_{yy}(\omega,k)=\sqrt{k^{2}-\omega^{2}}K^{T}(\omega,k).
\end{equation}
The expression~(\ref{KLKT}) can be seen as a consequence of electric-magnetic (EM) self-duality of Maxwell theory in four dimensional spacetime, which results in a frequency-independent conductivity.

When we consider higher derivative interactions between the metric and the gauge field in the bulk, the self-duality is broken and one may wonder what will happen in the dual CFT. One typical example of such kind of higher derivative interactions is given by
\begin{equation}
\label{CF2sec1}
S_{A}=\frac{1}{g_{4}^{2}}\int d^{4}x\sqrt{-g}[-\frac{1}{4}F_{\mu\nu}F^{\mu\nu}
+\gamma C_{\mu\nu\rho\sigma}F^{\mu\nu}F^{\rho\sigma}],
\end{equation}
where $C_{\mu\nu\rho\sigma}$ denotes the Weyl tensor. (\ref{CF2sec1}) has already appeared in previous holographic studies, e.g.~\cite{Hofman:2008ar, Hofman:2009ug, Ritz:2008kh}. The bounds on $\gamma$ were determined in~\cite{Hofman:2008ar, Hofman:2009ug} by imposing causality in the dual CFT and the conductivity was discussed in~\cite{Ritz:2008kh} with emphasis on five dimensional bulk spacetime. It was shown in~\cite{Myers:2010pk} that in four dimensional bulk spacetime, when $\gamma>0$ (the only parameter controlling the higher derivative corrections to the Maxwell theory), the conductivity exhibits similar form as that from the Boltzmann analysis: a peak at small frequency which is connected to the constant value at large frequency. When $\gamma<0$, it is the excitations of the EM dual theory that provide the Boltzmann-like interpretation of the conductivity. Moreover, it was shown in~\cite{Myers:2010pk} that even though the self-duality is lost, the following relation still holds,
\begin{equation}
\label{KLKHT}
K^{L}(\omega,k)\hat{K}^{T}(\omega,k)={\rm const},
\end{equation}
where $\hat{K}^{T}(\omega,k)$ denotes the counterpart in the EM dual theory. Subsequently the quasinormal modes (QNMs) in charge response function were studied in~\cite{Sachdev:2012prb, Sachdev:2013prb}, where the role of particle-vortex or S-duality was also considered. In particular, it was observed in~\cite{Sachdev:2012prb} that the holographic calculations are the first that satisfy both of the two sum rules originated from the quantum critical conductivity and the extension of the sum rules to finite momentum case was performed in~\cite{Sachdev:2013prb}. It should be emphasized that even though the small-frequency observed in~\cite{Myers:2010pk} has the same single-pole structure as the conventional Drude peak, the key difference is that the translation invariance is not broken in this case and the background is at zero charge density. Hence it is better to call the small-frequency peak Damle-Sachdev (DS) peak in the sense of~\cite{Damle:1997prb} rather than the `Drude' peak~\cite{Sachdev:2013prb}.

However, all the above mentioned results were obtained in the large $N$ limit, where the dual gravity background is Schwarzschild-${\rm AdS}_{4}$ black hole. Therefore it would be interesting to study the finite $N$ effects to the conductivity, in particular, to see if~(\ref{KLKT}) and~(\ref{KLKHT}) still hold at finite $N$. For concreteness in this paper we focus on one particular class of perturbative black hole solutions obtained in~\cite{Smolic:2013gz}, that is, black hole solutions in Einstein gravity plus $C^{3}$ corrections with $C$ being the Weyl tensor. We introduce a coupling constant $\alpha>0$ for the $C^{3}$ term. This kind of solutions can be seen as the analogue of Gauss-Bonnet black holes in four dimensional bulk spacetime in some sense~\cite{Smolic:2013gz}, hence provide natural candidates for investigations on finite $N$ effects in AdS/CMT. The effective action describing the charge transport is given by the standard Maxwell term plus an interacting term between the Weyl tensor and the gauge field strength, which was studied in~\cite{Myers:2010pk}. We calculate the DC conductivity using both the membrane paradigm and Kubo's formula, where the results precisely match. We compute the AC conductivity and find that in standard Maxwell theory, the AC conductivity is still frequency-independent even in the presence of $C^{3}$ corrections. The AC conductivity possesses quite distinct features when the interacting term is added. When $\gamma>0$, the amplitude of the DS peak increases as $\alpha$ or $\gamma$ grows bigger. When $\gamma<0$, the amplitude of the dip in the zero frequency limit may decrease to zero at certain particular value of $\alpha$ with fixed $\gamma$. Even though the background is at finite temperature and zero density, it is fair to say that the holographic calculations exhibit some analogue of the crossover from a `metal' to a `bad metal'. In both cases, the low frequency conductivity is connected smoothly to the constant value in the large frequency limit. In addition, we show that the relations~(\ref{KLKT}) and ~(\ref{KLKHT}) hold in this background. Actually the derivation for~(\ref{KLKHT}) does not explicitly depend on the concrete form of metric, which signifies its universality.

The rest of the paper is organized as follows: In section 2 we review the background obtained in~\cite{Smolic:2013gz}. The DC conductivity and the diffusion constant are calculated in section 3 via the membrane paradigm. In section 4 the DC conductivity is revisited using Kubo's formula, where the result precisely matches the one obtained in section 3. We explore the AC conductivity in section 5 and derive the relations~(\ref{KLKT}) and~(\ref{KLKHT}) in section 6. Finally a summary is given in section 7.
\section{The background}
We review the solutions obtained in~\cite{Smolic:2013gz} and their thermodynamics in this section. Gauss-Bonnet black holes in AdS
are useful backgrounds for studying finite $N$ effects in AdS/CFT in spacetime dimensions higher than four. However, the Gauss-Bonnet terms become trivial in four dimensional bulk spacetime, therefore in order to explore finite $N$ effects in AdS/CMT, one has to consider other types of higher derivative corrections. Various corrections to Einstein gravity in $AdS_{4}$ were discussed in~\cite{Smolic:2013gz}, including $R^{2}$, $f(R)$, $C^{3}$ and $R^{4}$ terms, where $R$ and $C$ denote the Riemann tensor and Weyl tensor respectively. Among all the perturbative black hole solutions, the one in Einstein$+C^{3}$ theory is a natural candidate for exploring finite $N$ effects in AdS/CMT. The reason is that even though the black hole solutions receive higher order corrections, there are no new operators induced in the spectrum, hence they are in some sense analogous to Gauss-Bonnet black holes in higher dimensions.\footnote{For details see~\cite{Smolic:2013gz}.}

We start from the following effective action,
\begin{equation}
\label{action}
S=\frac{1}{2\kappa^{2}}\int d^{4}x\sqrt{-g}[R-2\Lambda+
\alpha{C_{\mu\nu}}^{\rho\sigma}{C_{\rho\sigma}}^{\eta\lambda}{C_{\eta\lambda}}^{\mu\nu}],
\end{equation}
where $C_{\mu\nu\rho\sigma}$ is the Weyl tensor and $\alpha$ is the corresponding coupling constant. Here we take $\alpha>0$ without loss of generality. We fix the cosmological constant $\Lambda=-3$ and take the ansatz for the metric as follows
\begin{equation}
\label{ansatz}
ds^{2}=-a(r)b(r)^{2}dt^{2}+\frac{dr^{2}}{a(r)}+r^{2}(dx^{2}+dy^{2}).
\end{equation}
Generically the equations of motion are complicated and we have to solve for the metric perturbatively at the leading order of $\alpha$. The simplest way to work out the corrections to the planar AdS black holes is to substitute the ansatz~(\ref{ansatz}) into the action~(\ref{action}) and reduce the original action to a one-dimensional effective action. Then we can obtain the equations of motion for $a(r)$ and $b(r)$ by varying the one-dimensional action and solve for them perturbatively in $\alpha$. The solutions for $a(r)$ and $b(r)$ to order $\alpha$ are given by
\begin{eqnarray}
a(r)&=&a_{0}(r)+\alpha a_{1}(r)\nonumber\\
&=&r^{2}-\frac{m}{r}+\alpha(\frac{a_{1}}{r}+\frac{24m^{2}}{r^{4}}-\frac{16m^{3}}{r^{7}}),\nonumber\\
b(r)&=&b_{0}(r)+\alpha b_{1}(r)=1+\alpha(b_{1}-\frac{6m^{2}}{r^{6}}),
\end{eqnarray}
where $a_{1}$ and $b_{1}$ are arbitrary integration constants. Note that $a_{1}$ induces a redefinition of the mass parameter $m$ at order $\alpha$ as
$m\rightarrow m-\alpha a_{1}$ and $b_{1}$ changes the norm of the timelike Killing vector at infinity at order $\alpha$. In~\cite{Smolic:2013gz} $a_{1}$ and $b_{1}$ were set to zero.

Next we discuss the thermodynamics of the corrected black hole solutions. When $a_{1}=b_{1}=0$, the horizon $r_{H}$, determined by $a(r_{H})=0$,
is given by the following result to order $\alpha$,
\begin{eqnarray}
& &r_{H}=r_{H(0)}+\alpha r_{H(1)},\nonumber\\
& &r_{H(0)}^{3}=m,~~~r_{H(1)}=-\frac{8}{3}r_{H(0)}.
\end{eqnarray}
Then the temperature is
\begin{equation}
\label{T0}
T=\frac{a^{\prime}(r_{H})b(r_{H})}{4\pi}=\frac{3}{4\pi}r_{H(0)}(1-\frac{2}{3}\alpha).
\end{equation}
The black hole entropy can be obtained via the Wald formula, which leads to
\begin{equation}
\label{S0}
S=\frac{r_{H(0)}^{2}}{4}(1+\frac{2}{3}\alpha).
\end{equation}
It can be seen that even though both $T$ and $S$ are corrected at order $\alpha$, the combination $TS$ is uncorrected
at this order. Then upon imposing the first law of thermodynamics $dM=TdS$, the mass should also be unchanged
at order $\alpha$, so is the free energy. These arguments have been confirmed by straightforward calculations in~\cite{Smolic:2013gz}.

For later convenience, we require that $r_{H(0)}$ is still the horizon, which gives $a_{1}=-8r_{H(0)}^{3}$. Then the black hole metric can be rewritten as
\begin{eqnarray}
\label{bhrzb}
ds^{2}&=&-r^{2}f(r)b(r)^{2}dt^{2}+\frac{dr^{2}}{r^{2}f(r)}+r^{2}(dx^{2}+dy^{2}),~f(r)=f_{0}(r)+\alpha f_{1}(r),\nonumber\\
f_{0}(r)&=&1-\frac{r_{H(0)}^{3}}{r^{3}},~~~f_{1}(r)=24\frac{r_{H(0)}^{6}}{r^{6}}-8\frac{r_{H(0)}^{3}}{r^{3}}-16\frac{r_{H(0)}^{9}}{r^{9}},\nonumber\\
b(r)&=&1-6\alpha\frac{r_{H(0)}^{6}}{r^{6}}.
\end{eqnarray}
Introducing the radial coordinate $u=r_{H(0)}/r$, the black hole solution is given by
\begin{eqnarray}
\label{bhuzb}
ds^{2}&=&-\frac{r_{H(0)}^{2}}{u^{2}}f(u)b(u)^{2}dt^{2}+\frac{du^{2}}{u^{2}f(u)}+\frac{r_{H(0)}^{2}}{u^{2}}(dx^{2}+dy^{2}),~~f(u)=f_{0}(u)+\alpha f_{1}(u),\nonumber\\
f_{0}(u)&=&1-u^{3},~~~f_{1}(u)=24u^{6}-8u^{3}-16u^{9},\nonumber\\
b(u)&=&1-6\alpha u^{6},
\end{eqnarray}
where the horizon is located at $u=1$. Moreover, in order to avoid the undesirable zeros in $b(u)$, we impose the constraint $\alpha<1/6$. The temperature is
\begin{equation}
\label{tempsec2}
T=\frac{3}{4\pi}r_{0}(1+2\alpha),
\end{equation}
where we denote $r_{0}\equiv r_{H(0)}$ for simplicity.
The thermodynamics of the black holes with $a_{1}\neq0$ can be discussed in a simpler way. Recall that the mass parameter shifts as
$m\rightarrow m-\alpha a_{1}$ in the presence of a nontrivial $a_{1}$. So the temperature and the entropy shift as
\begin{equation}
T\rightarrow T(1+2\alpha),~~~S\rightarrow S(1+6\alpha),
\end{equation}
where $T$ and $S$ on the RHS denote the values in~(\ref{T0}) and~(\ref{S0}). Moreover, the free energy and the energy shift as
\begin{equation}
F\rightarrow F(1+8\alpha),~~~M\rightarrow M(1+8\alpha).
\end{equation}
Therefore the first law of thermodynamics still holds at the leading order of $\alpha$. Note that since the perturbative metric is valid up to the first order of $\alpha$, all the calculations in the following sections are also performed to the same order.

\section{Conductivity from the membrane paradigm}
We calculate the DC conductivity in our perturbative black hole background by using the membrane paradigm.
We consider the following action for the $U(1)$ gauge field, following~\cite{Myers:2010pk},
\begin{equation}
\label{CF2}
S_{A}=\frac{1}{g_{4}^{2}}\int d^{4}x\sqrt{-g}[-\frac{1}{4}F_{\mu\nu}F^{\mu\nu}
+\gamma C_{\mu\nu\rho\sigma}F^{\mu\nu}F^{\rho\sigma}].
\end{equation}
The corresponding equation of motion is given by
\begin{equation}
\partial_{\mu}[\sqrt{-g}(F^{\mu\nu}-4\gamma C^{\mu\nu\rho\sigma}F_{\rho\sigma})]=0.
\end{equation}
The first term in~(\ref{CF2}) is the ordinary Maxwell term and the second one gives an interacting term between the Weyl tensor and the field strength. As claimed in~\cite{Myers:2010pk}, this particular interaction has the advantage that the charge transport at zero temperature remains unchanged, as the Weyl curvature vanishes in pure AdS geometry. The calculations in~\cite{Myers:2010pk} were performed in Schwarzschild-${\rm AdS}_{4}$ black hole background with $N\rightarrow\infty$ and here we will see how the DC conductivity behaves at finite $N$.

In order to apply the membrane paradigm, we first rewrite the action~(\ref{CF2}) as
\begin{equation}
S_{A}=\int d^{4}x\sqrt{-g}(-\frac{1}{8g_{4}^{2}}F_{\mu\nu}X^{\mu\nu\rho\sigma}F_{\rho\sigma}),
\end{equation}
where $X^{\mu\nu\rho\sigma}$ satisfies
\begin{equation}
X^{\mu\nu\rho\sigma}=X^{[\mu\nu][\rho\sigma]}=X^{\rho\sigma\mu\nu}.
\end{equation}
When the action is given by the Maxwell term, $X$ is simply
\begin{equation}
\label{Xmax}
{X_{\mu\nu}}^{\rho\sigma}={I_{\mu\nu}}^{\rho\sigma}={\delta_{\mu}}^{\rho}{\delta_{\nu}}^{\sigma}
-{\delta_{\mu}}^{\sigma}{\delta_{\nu}}^{\rho},
\end{equation}
while $X$ is given by the following expression in the presence of Weyl coupling
\begin{equation}
\label{Xweyl}
{X_{\mu\nu}}^{\rho\sigma}={I_{\mu\nu}}^{\rho\sigma}-8\gamma {C_{\mu\nu}}^{\rho\sigma}.
\end{equation}
Given these results, it is straightforward to proceed using the membrane paradigm~\cite{Iqbal:2008by}. The associated current reads
\begin{equation}
j^{\mu}=\frac{1}{4}n_{\nu}X^{\mu\nu\rho\sigma}F_{\rho\sigma}\Big|_{r=r_{H}},
\end{equation}
where $n_{\nu}$ is an outward-pointing radial unit vector. Hence the diffusion constant is given by
\begin{equation}
D=-\sqrt{-g}\sqrt{-X^{xtxt}X^{xrxr}}\Big|_{r=r_{H}}\int^{\infty}_{r_{H}}\frac{dr}{\sqrt{-g}X^{trtr}}.
\end{equation}
According to Ohm's law, the DC conductivity is
\begin{equation}
\sigma_{DC}=\frac{1}{g_{4}^{2}}\sqrt{-g}\sqrt{-X^{xtxt}X^{xrxr}}\Big|_{r=r_{H}}.
\end{equation}
Finally the susceptibility $\chi$ is easily determined by the Einstein relation $D=\sigma_{DC}/\chi$
\begin{equation}
\chi^{-1}=-\int^{\infty}_{r_{H}}\frac{dr}{\sqrt{-g}X^{trtr}}.
\end{equation}

First let us consider the Maxwell term, where $X$ is given by~(\ref{Xmax}). The diffusion constant, the DC conductivity and the susceptibility
are
\begin{equation}
D=\frac{1}{r_{H(0)}}(1-\frac{6}{7}\alpha)=\frac{3}{4\pi T}(1+\frac{8}{7}\alpha),
\end{equation}
\begin{equation}
\sigma_{DC}=\frac{1}{g_{4}^{2}},
\end{equation}
\begin{equation}
\chi=\frac{4\pi T}{3g_{4}^{2}}(1-\frac{8}{7}\alpha).
\end{equation}
Note that the DC conductivity is still a constant, and does not receive any $\alpha$ corrections. This may be due to the electro-magnetic duality in four dimensional bulk spacetime. Later we will see that the AC conductivity is still a frequency-independent constant even with $C^{3}$ corrections.
Once the Weyl tensor is included, the DC conductivity reads
\begin{equation}
\label{ddcsec3}
\sigma_{DC}=\frac{1}{g_{4}^{2}}[1+4\gamma(1+26\alpha)].
\end{equation}
It can be seen that when $\alpha=0$, the result reduces to that obtained in~\cite{Myers:2010pk}.

\section{Conductivity from the Green's function}
In this section we calculate the DC conductivity in the presence of Weyl coupling via Kubo's formula. Since we are interested in the DC conductivity, it is sufficient to set the momentum $k=0$. We get rid of the factor $e^{-i\omega t}$ from the Fourier transformation mode and still denote the new variable as $A_{y}$ for notational simplicity. The relevant equation of motion reads
\begin{equation}
A_{y}^{\prime\prime}+\frac{h^{\prime}}{h}A^{\prime}_{y}+\omega^{2}PA_{y}=0,
\end{equation}
where $h(u)$ and $P(u)$ are given by
\begin{eqnarray}
h &=&\sqrt{-g}g^{uu}g^{yy}(1-8\gamma g^{uu}g^{yy}C_{uyuy})\nonumber\\
&=&r_{0}bf\left(1-\frac{2\gamma u^{2}}{3b}(3b^{\prime}f^{\prime}+2fb^{\prime\prime}+bf^{\prime\prime})\right)\nonumber\\
&\equiv&r_{0}bfK,\nonumber\\
P &=&-\frac{g^{tt}(1-8\gamma g^{tt}g^{yy}C_{tyty})}{g^{uu}(1-8\gamma g^{uu}g^{yy}C_{uyuy})}\nonumber\\
&=&\frac{1}{r_{0}^{2}b^{2}f^{2}},
\end{eqnarray}
We will extract the retarded Green's function via AdS/CFT, following~\cite{Son:2002sd}. First we expand the action as
\begin{equation}
S_{A_{y}}=-\frac{1}{2g_{4}^{2}}\int d^{4}xhA_{y}^{\prime2},
\end{equation}
which results in the following retarded Green's function
\begin{equation}
G^{R}_{yy}=-\frac{1}{g_{4}^{2}}h\frac{A_{y}^{\prime}}{A_{y}}\Big|_{u\rightarrow0}.
\end{equation}

To obtain the retarded Green's function, in principle we should solve the corresponding equation of motion in a hydrodynamic expansion.
However, as pointed out in~\cite{CaronHuot:2006te, Atmaja:2008mt}, there exists a short cut for calculating the DC conductivity. For a general second order differential equation
\begin{equation}
\label{eomaysec4}
A_{y}^{\prime\prime}+\frac{h^{\prime}}{h}A^{\prime}_{y}+\omega^{2}PA_{y}=0,
\end{equation}
there is a conserved quantity
\begin{eqnarray}
Q &\equiv&e^{\int\frac{h^{\prime}}{h}}(\bar{A}_{y}\partial_{u}A_{y}-A_{y}\partial_{u}\bar{A}_{y})\nonumber\\
&=&h(\bar{A}_{y}\partial_{u}A_{y}-A_{y}\partial_{u}\bar{A}_{y}).
\end{eqnarray}
Then we can reexpress the retarded Green's function in terms of the conserved quantity as
\begin{equation}
{\rm Im}G^{R}_{yy}=-\frac{1}{2ig_{4}^{2}}\frac{Q}{|A_{y}(0)|^{2}}.
\end{equation}
As shown in~\cite{Son:2002sd}, $A_{y}$ admits the following near-horizon expansion
\begin{equation}
A_{y}(u)=(u_{H}-u)^{-\frac{i\omega}{4\pi T}}y(u),
\end{equation}
Since $Q$ is a conserved quantity, we can evaluate it at the horizon $u=u_{H}$, which gives
\begin{equation}
Q=-2i\omega|y(u_{H})|^{2}K(u_{H}).
\end{equation}
The retarded Green's function is therefore
\begin{equation}
{\rm Im}G^{R}_{yy}=\frac{\omega}{g_{4}^{2}}K(u_{H})\frac{|y(u_{H})|^{2}}{|y(0)|^{2}}.
\end{equation}
Note that in the low frequency limit, the solution to the equation of motion~(\ref{eomaysec4}) can be well approximated as $y(u)={\rm const}$, so
\begin{equation}
{\rm Im}G^{R}_{yy}=\frac{\omega}{g_{4}^{2}}K(u_{H}).
\end{equation}
Finally the DC conductivity reads
\begin{eqnarray}
\sigma_{DC}&=&\lim_{\omega\rightarrow0}\frac{{\rm Im}G^{R}_{yy}}{\omega}\nonumber\\
&=&\frac{K(u_{H})}{g_{4}^{2}}=\frac{1}{g_{4}^{2}}[1+4\gamma(1+26\alpha)],
\end{eqnarray}
which agrees with that obtained via the membrane paradigm.
\section{The AC conductivity}
We study the AC conductivity in this section. First for comparison with the results in condensed matter physics, we briefly review the conductivity of a CFT at finite temperature. Then we turn to the holographic calculations of the AC conductivity and consider both the standard Maxwell theory and the $CF^{2}$ corrections. We find that in the standard Maxwell theory, the conductivity is still a frequency-independent constant, which is the same as the $N\rightarrow\infty$ case. When the $CF^{2}$ corrections are taken into account, we observe a Drude-like peak in the $\omega\rightarrow0$ limit when $\gamma>0$. As suggested in~\cite{Sachdev:2013prb}, it would be proper to call such a peak the Damle-Sachdev (DS) peak~\cite{Damle:1997prb} rather than the Drude peak in the usual sense. When $\gamma<0$, the amplitude of the low-frequency dip becomes very small at certain particular value of $\alpha$. The corresponding AC conductivity exhibits the typical behavior of an insulator. Even though the background is at finite temperature and zero charge density, it is fair to say that our result provides certain analogue of the crossover from `metal' to `bad metal'. In both cases, the low frequency conductivity is connected smoothly to the constant value in the large frequency limit.
\subsection{The conductivity of a CFT at finite temperature}
The charge transport properties of systems near a quantum critical point have been a long-standing problem in both experimental and theoretical condensed matter physics. As emphasized in~\cite{Damle:1997prb}, there exists an important distinction between experimental and theoretical studies: The experiments are carried out at a low temperature $T\neq0$ and the DC conductivity is measured at frequency $\omega$ which satisfies $\hbar\omega\ll k_{B}T$, while the theoretical analysis focuses on the regime $\hbar\omega\gg k_{B}T$. Indeed the behavior of charge transport in two qualitatively different regimes is given by
\begin{itemize}
\item $\hbar\omega\ll k_{B}T$: The hydrodynamic, incoherent, collision-dominated regime; the conductivity should exhibit a Drude-like peak as a function of frequency.
\item $\hbar\omega\gg k_{B}T$: The high-frequency, phase-coherent, collisionless regime; the transport is dominated by the external perturbation.
\end{itemize}
Based on the above facts, the authors of~\cite{Damle:1997prb} proposed the following formula for the conductivity in $d$ spatial dimensions,
\begin{equation}
\sigma(\omega)=\frac{Q^{2}}{\hbar}\left(\frac{k_{B}T}{\hbar c}\right)^{(d-2)/z}\Sigma\left(\tilde{\omega}\right),~~\tilde{\omega}\equiv\frac{\hbar\omega}{k_{B}T},
\end{equation}
where $Q$ denotes the `charge' carried by the order parameter quanta, $z$ is the dynamical exponent, $c$ is a non-universal microscopically determined quantity with dimension $[{\rm length}]^{z}[{\rm time}]^{-1}$ and $\Sigma(\tilde{\omega})$ is a universal scaling function. $\Sigma(\tilde{\omega})$ possesses the following characteristic features:
\begin{itemize}
\item The DC conductivity is determined by $\Sigma(0)$;
\item At small $\tilde{\omega}$ there is a Drude-like peak due to the energy-exchanging collisions among thermally excited carriers;
\item As $\tilde{\omega}$ increases there is crossover to transport determined by the external source;
\item As $\tilde{\omega}\rightarrow\infty$, it is expected that $\Sigma(\tilde{\omega})\sim(-i\tilde{\omega})^{(d-2)/z}$. When $d=2$, $\Sigma(\infty)$ is a real, finite, universal number determining the high frequency conductivity.
\end{itemize}
We will show that these features can be reproduced in the holographic calculations in the next subsections.
\subsection{The holographic AC conductivity}
In this subsection we evaluate the AC conductivity in the perturbative black hole background. The action for the $U(1)$ gauge field is the standard Maxwell term plus an interacting term between the Weyl tensor and the field strength, given by~(\ref{CF2}). The equation of motion is given by
$$\partial_{\mu}[\sqrt{-g}(F^{\mu\nu}-4\gamma C^{\mu\nu\rho\sigma}F_{\rho\sigma})]=0.$$ The Fourier expansion of the gauge field fluctuations can be written as
\begin{equation}
A_{\mu}(t,x,u)=\int\frac{d^{3}k}{(2\pi)^{3}}e^{-i\omega t+ikx}A_{\mu}(u),
\end{equation}
and we work in the gauge $A_{u}=0$. The corresponding equation of motion for $A_{y}$ is
\begin{equation}
A_{y}^{\prime\prime}+\frac{h^{\prime}}{h}A^{\prime}_{y}+(\omega^{2}P_{1}+k^{2}P_{2})A_{y}=0,
\end{equation}
where $h(u), P_{1}(u)$ and $P_{2}(u)$ are given by
\begin{eqnarray}
h&=&\sqrt{-g}g^{uu}g^{yy}(1-8\gamma g^{uu}g^{yy}C_{uyuy})\nonumber\\
&=&r_{0}bf\left(1-\frac{2\gamma u^{2}}{3b}(3b^{\prime}f^{\prime}+2fb^{\prime\prime}+bf^{\prime\prime})\right),\nonumber\\
P_{1}&=&-\frac{g^{tt}(1-8\gamma g^{tt}g^{yy}C_{tyty})}{g^{uu}(1-8\gamma g^{uu}g^{yy}C_{uyuy})}\nonumber\\
&=&\frac{1}{r_{0}^{2}b^{2}f^{2}},\\
P_{2}&=&-\frac{g^{xx}(1-8\gamma g^{xx}g^{yy}C_{xyxy})}{g^{uu}(1-8\gamma g^{uu}g^{yy}C_{uyuy})}\nonumber\\
&=&\frac{4\gamma u^{2}\left(3b^{\prime}f^{\prime}+2fb^{\prime\prime}+b(3+4\gamma u^{2}f^{\prime\prime})\right)}
{r_{0}^{2}f\left(2\gamma u^{2}(3b^{\prime}f^{\prime}+2fb^{\prime\prime})+b(-3+2\gamma u^{2}f^{\prime\prime})\right)},\nonumber
\end{eqnarray}
Since we are concerning about the frequency-dependent conductivity, it is sufficient to take $k=0$. The conductivity is given by
\begin{equation}
\label{condsec5}
\sigma(\omega)=-{\rm Im}\left(\frac{G^{R}_{yy}}{\omega}\right),
\end{equation}
where the retarded Green's function $G^{R}_{yy}$ reads
\begin{equation}
G^{R}_{yy}=-\frac{1}{g_{4}^{2}}h\frac{A_{y}^{\prime}}{A_{y}}\Big|_{u\rightarrow0}.
\end{equation}
In the following we will work in the dimensionless quantity $\bar{\omega}=\omega/r_{0}$ and set $g_{4}=1$ for convenience.

Due to its complexity of the corresponding equation of motion, we have to solve the equation numerically. We apply the Frobenius method to approximate the solution at the horizon by series expansion and use the Runge-Kutta method to solve the initial condition problem. The general solution can be expressed as
\begin{equation}
A_{y}(u)=(1-u)^{\beta}F(u),~~~\beta=-\frac{i\omega}{4\pi T},
\end{equation}
where $T$ denotes the temperature given by~(\ref{tempsec2}). We need to fix our initial conditions at the horizon $u=1$ and numerically integrate the equation. Once the numerical solution is obtained, the conductivity can be evaluated by~(\ref{condsec5}). Now we have two parameters, $\alpha$ characterizing the effect of the higher derivative corrections and $\gamma$ characterizing the interaction between the Weyl tensor and the gauge field strength. For nonvanishing $\alpha$, the background metric would differ from the original Schwarzschild-${\rm AdS}_{4}$ background and it is expected that we could observe nontrivial behavior for the conductivity, compared to that obtained in~\cite{Myers:2010pk}.

We can fix the bounds on $\gamma$ by requiring that the dual CFT should preserve causality, following~\cite{Myers:2010pk}. The detailed analysis will be presented in Appendix A and the bounds turn out to be
\begin{equation}
-\frac{1}{12}\leq\gamma\leq\frac{1}{12},
\end{equation}
which is the same as the result in~\cite{Myers:2010pk}. In the following cases we consider $\alpha\in[0, 0.06]$.


\underline{A. $\gamma>0$}
\begin{figure}
\begin{center}
\vspace{-1cm}
\hspace{-0.5cm}
\includegraphics[angle=0,width=0.45\textwidth]{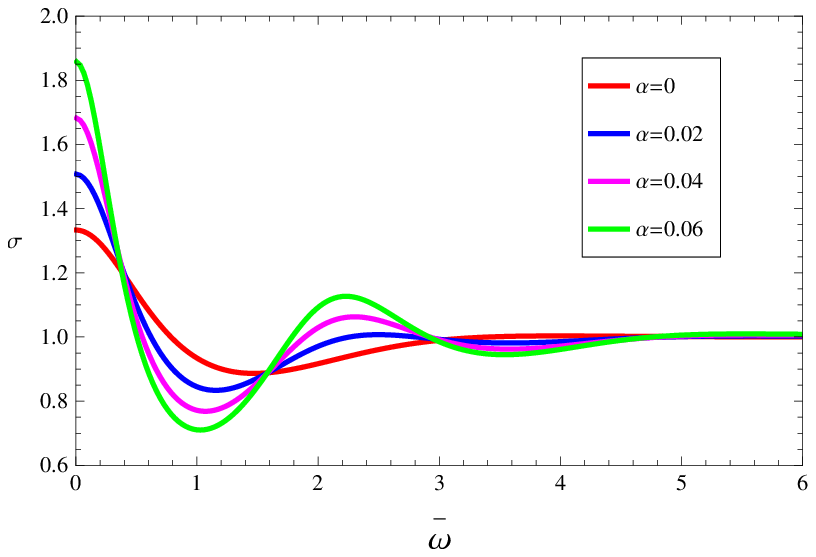}
\includegraphics[angle=0,width=0.45\textwidth]{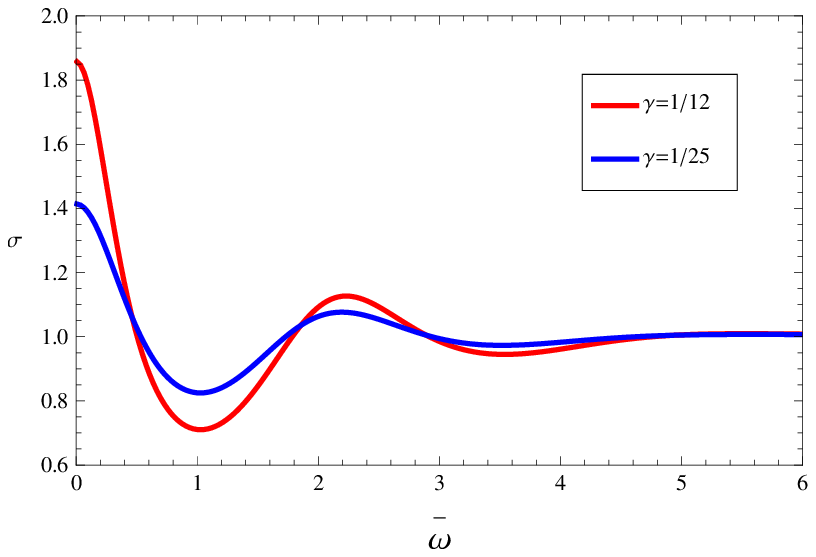}
\caption{\small Left: The AC conductivity in the range $\bar{\omega}\in[0,6]$ with different $\alpha$ at $\gamma=1/12$.
Right: Comparison of the AC conductivity with $\alpha=0.06$ and $\gamma=1/25, 1/12$ respectively.}
\label{gamma008}
\end{center}
\end{figure}

We plot the AC conductivity $\sigma(\bar{\omega})$ with $\gamma=1/12$ and different $\alpha$ in Fig.~\ref{gamma008}. Here are some remarks on the results.
\begin{itemize}
\item It can be seen in the left plot that when $\alpha=0$, we can reproduce the result obtained in~\cite{Myers:2010pk}. The DS peak appears at $\bar{\omega}=0$. 
\item As $\alpha$ increases, the AC conductivity develops a second peak at certain intermediate frequency $\bar{\omega}\approx2.10$.
\item The right plot shows the behavior of the AC conductivity with different $\gamma$ at fixed $\alpha$. Here we focus on the case $\alpha=0.06$. The amplitude of the DS peak increases and the second peak becomes more apparent as $\gamma$ grows larger.
\item The AC conductivity becomes constant in the large frequency limit, irrespective of the values of $\gamma$ or $\alpha$. This can be seen more explicitly in the right plot.  It can be seen that when $\bar{\omega}$ becomes large, the conductivity tends to constant (normalized to be one here).
\end{itemize}

\underline{B. $\gamma<0$}

\begin{figure}
\begin{center}
\vspace{1cm}
\hspace{0.5cm}
\includegraphics[angle=0,width=0.45\textwidth]{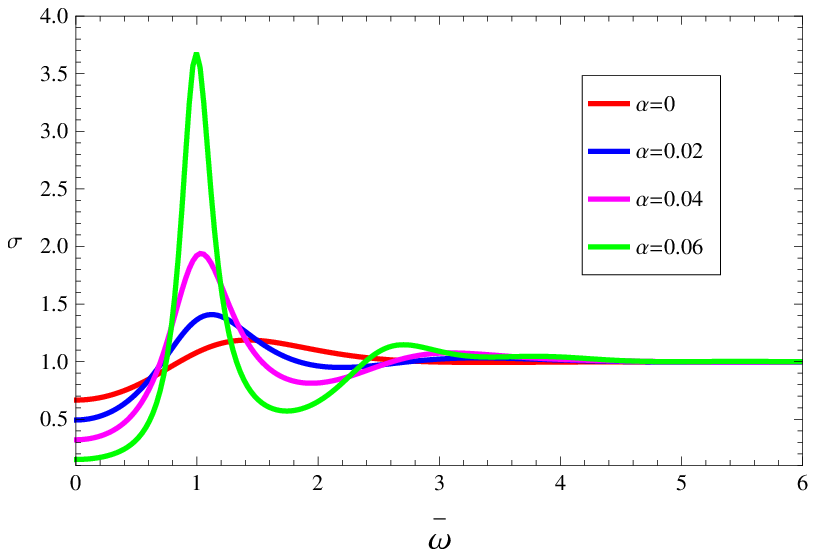}
\includegraphics[angle=0,width=0.45\textwidth]{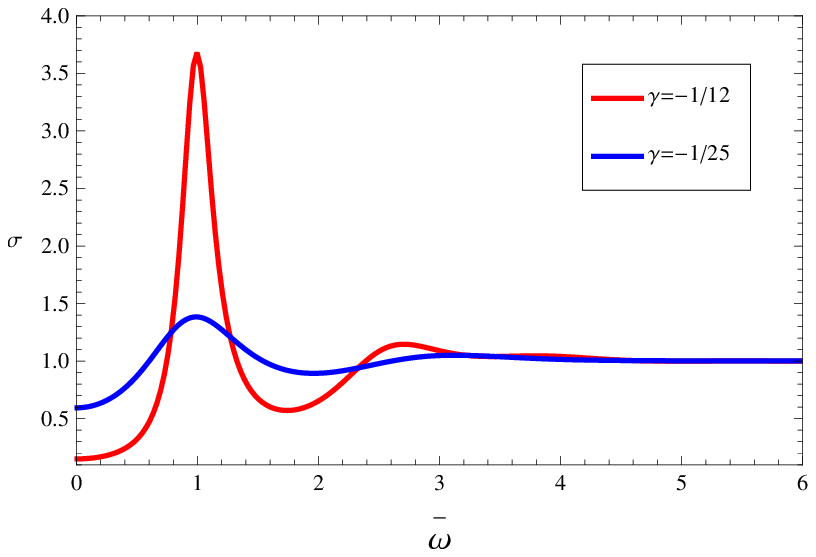}
\caption{\small Left: The AC conductivity at different $\alpha$ with $\gamma=-1/12$. Right: Comparison of the AC conductivity with $\alpha=0.06$ and $\gamma=-1/25, -1/12$ respectively.}
\label{gammamin00823}
\end{center}
\end{figure}

Now we turn to the lower bound of $\gamma$ and plot the conductivity. The results are shown in Fig.~\ref{gammamin00823} and we can observe the following features:
\begin{itemize}
\item The minimum value of the conductivity appears at $\bar{\omega}=0$ and it approaches zero at some particular value of $\alpha$, which can be approximately obtained from~(\ref{ddcsec3}). Moreover the zero DC conductivity cannot be always achieved for smaller $|\gamma|$ because $\alpha$ could go beyond the physical regime.
\item The behavior of the AC conductivity looks analogous to the insulator when the DC conductivity can take zero value, even though the system is at finite temperature and zero density.
\item As $\alpha$ increases, the AC conductivity develops a second peak around $\bar{\omega}\approx1.0$. The amplitude of the peak increases as $\alpha$ grows bigger.
\item In the right plot it can be seen that with the same $\alpha$, the DC conductivity decreases and the amplitude of the second peak increases when $\gamma$ becomes smaller.
\item The conductivity tends to be a constant in the large frequency limit, which is the same as the $\gamma>0$ case.
\end{itemize}
Before moving on, let us consider some fundamental aspects of metal-insulator transition. Metal-insulator transition driven by electron interactions is a key feature in the phase diagram of several classes of materials, e.g. the cuprate superconductors~\cite{Imada:1998rmp, Dobro:2011oup}. The DC conductivity drops dramatically as the system transitions to an insulating phase, while the Drude peak in the optical conductivity becomes a gap in the low energy spectral weight. One can encounter `bad metals' in the vicinity of the metal-insulator transition, characterized by metallic behavior in the absence of a Drude peak and have resistivities exceeding the Mott-Ioffe-Regel limit~\cite{Gun:2003rmp, Hus:2004phi}. In our holographic setup, the DS peak can be flattened as positive $\gamma$ decreases and the DC conductivity decreases when $\gamma$ approaches the lower bound $-1/12$ or $\alpha$ is bigger. It may be zero at certain $\alpha$ with fixed negative $\gamma$. This behavior may be seen as an analog of the crossover from a `metal' to a `bad' metal. It should be emphasized that the corresponding background may not be considered as a holographic insulating phase because it is at finite temperature and zero density.


\underline{C. $\gamma=0$}

The conductivity is still constant, which can be normalized to be one. This is the same as the result obtained in~\cite{Herzog:2007ij}, where the conductivity was calculated in the large $N$ limit. It may indicate that the universal relation~(\ref{KLKT}) observed in~\cite{Herzog:2007ij} still holds at finite $N$, at least in our particular background. We will revisit this topic in the next section.

\section{Universal relations for the Green's functions}
The numerical result that when $\gamma=0$, the conductivity is still a frequency-independent constant indicates that the universal relation for retarded Green's functions obtained in~\cite{Herzog:2007ij} may still hold. In this section we calculate the retarded Green's functions and find that the universal relations for the Green's functions~(\ref{KLKT}) and~(\ref{KLKHT}) still hold in our background with higher derivative corrections.

Generically for a CFT at finite temperature $T>0$, current conservation and spatial rotational invariance fix the
following general structure of the retarded Green's functions
\begin{equation}
G^{R}_{\mu\nu}({\bf k})=\sqrt{{\bf
k^{2}}}(P^{T}_{\mu\nu}K^{T}(\omega,k)+P^{L}_{\mu\nu}K^{L}(\omega,k)),
\end{equation}
where ${\bf k}^{\mu}=(\omega,k^{x},k^{y}),
k^{2}=[(k^{x})^{2}+(k^{y})^{2}]^{1/2}, {\bf k}^{2}=k^{2}-\omega^{2}$.
Here $P^{T}_{\mu\nu}$ and $P^{L}_{\mu\nu}$ are orthogonal projection
operators given by
\begin{equation}
P^{T}_{tt}=P^{T}_{ti}=P^{T}_{it}=0,~~~P^{T}_{ij}=
\delta_{ij}-\frac{k_{i}k_{j}}{k^{2}},~~~
P^{L}_{\mu\nu}=\left(\eta_{\mu\nu}-
\frac{k_{\mu}k_{\nu}}{|{\bf k}|^{2}}\right)-P^{T}_{\mu\nu},
\end{equation}
where $i,j$ are spatial indices and $\mu,\nu$ denote the whole
spacetime indices. Let us choose ${\bf k}^{\mu}=(\omega,k,0)$ for
simplicity, then we have
\begin{equation}
\label{GRK}
G^{R}_{yy}(\omega,k)=\sqrt{k^{2}-\omega^{2}}K^{T}(\omega,k),~~~
G^{R}_{tt}(\omega,k)=-\frac{k^{2}}{\sqrt{k^{2}-\omega^{2}}}K^{L}(\omega,k).
\end{equation}
It was observed in~\cite{Herzog:2007ij} that at the leading order
level, i.e.  in the standard four-dimensional Maxwell theory and infinite $N$, $K^{T}$
and $K^{L}$ satisfy the following simple relation
$$K^{T}(\omega,k)K^{L}(\omega,k)={\rm const},$$ which signifies
self-duality of the theory. As a result, the conductivity is a fixed
constant. Later the authors of~\cite{Myers:2010pk} considered the Weyl corrections to the Maxwell action
and observed that even though the EM self-duality is broken, there exists a similar relation
$K^{L}(\omega,k)\hat{K}^{T}(\omega,k)={\rm const}$, where $\hat{K}^{T}(\omega,k)$ denotes the counterpart in the EM dual theory.
We will see that those equalities still hold in our background, which suggests that those relations do not receive finite $N$ corrections.

\subsection{The Maxwell theory}
First let us consider the Maxwell theory, where the equation of motion can be written as follows in components,
\begin{eqnarray}
\label{Maxequ}
& &A_{t}^{\prime\prime}-\frac{b^{\prime}}{b}A_{t}^{\prime}-\frac{1}{f}(\bar{\omega}qA_{x}+q^{2}A_{t})=0,\nonumber\\
& &\bar{\omega}A_{t}^{\prime}+qfb^{2}A_{x}^{\prime}=0,\nonumber\\
& &A_{x}^{\prime\prime}+(\frac{b^{\prime}}{b}
+\frac{f^{\prime}}{f})A_{x}^{\prime}+\frac{1}{b^{2}f^{2}}(\bar{\omega}^{2}A_{x}+\bar{\omega}qA_{t})=0,\nonumber\\
& &A_{y}^{\prime\prime}+(\frac{b^{\prime}}{b}
+\frac{f^{\prime}}{f})A_{y}^{\prime}+\frac{1}{b^{2}f^{2}}(\bar{\omega}^{2}-q^{2}b^{2}f)A_{y}=0.
\end{eqnarray}
Here $\bar{\omega}=\omega/r_{0},~q=k/r_{0}$.
We can obtain the following third order differential equation for $A_{t}$
\begin{equation}
A_{t}^{\prime\prime\prime}+(\frac{f^{\prime}}{f}-\frac{b^{\prime}}{b})A_{t}^{\prime\prime}
+\frac{1}{b^{2}f^{2}}(\bar{\omega}^{2}-q^{2}b^{2}f+b^{\prime2}f^{2}
-f^{\prime}fb^{\prime}b-b^{\prime\prime}f^{2}b)A_{t}^{\prime}=0.
\end{equation}
It can be seen that $A_{t}^{\prime}$ and $A_{y}$ do not have the same form of the equations of motion any more, in contrary to the case
with infinite $N$ studied in~\cite{Herzog:2007ij}.

The first step is to solve the following two equations of $A_{t}$
\begin{equation}
\label{at2p}
A_t'' - \frac{b'}{b} A_t' - \frac{1}{f} ( \bar{\omega} q A_x + q^2 A_t) =0,
\end{equation}
\begin{equation}
\label{at3p}
A_t ''' + ( \frac{f'}{f} - \frac{b'}{b} ) A_t '' + \frac{1}{b^2 f^2} ( \bar{\omega}^2- q^2 b^2+b'^2f^2 - f'fb'b - b''f^2 b)A_t '=0.
\end{equation}
We perform the series expansions for $A_{t}$ and $A_{x}$ as $A_t = \sum_{k=0}^\infty a_k u^k$, $A_x = \sum_{k=0}^\infty b_k u^k$. The strategy is to
solve~(\ref{at2p}) first and express the coefficients $a_k$ of $A_t$ in terms of $b_k$. Then we substitute the result into~(\ref{at3p}) and eliminate $b_k$. The final result is given by
\begin{eqnarray}
\label{At58}
A_t &= &a_0[ \  1 + \frac{a_1}{a_0} u + \frac{1}{2} q^2  u ^2 + \frac{1}{6} ( q^2-\bar{\omega}^2)  \frac{a_1}{a_0} u^3 + \frac{1}{24} q^2(q^2-\bar{\omega}^2)  u^4 + o(u^5)\ ]\nonumber\\
& &+ b_0 [ \ \frac{1}{2} q \omega u^2 + \frac{1}{24} ( q^3\bar{\omega} - q \bar{\omega}^3) u^4 + o(u^5)\ ].
\end{eqnarray}
Next we assume $A_\mu = M_{\mu\nu} A_\nu(0)$, which will lead to the relation $A_t =  M_{tt} A_t(0)+  M_{tx}A_x(0) $. Comparing with~(\ref{At58}) we arrive at
\begin{eqnarray}
M_{tt}&=& 1 + \frac{a_1}{a_0} u + \frac{1}{2} q^2  u ^2 + O(u^3),\nonumber\\
M_{tx }&=&\frac{1}{2} q \bar{\omega} u^2 + \frac{1}{24} ( q^3 \bar{\omega} - q \bar{\omega}^3) u^4 + o(u^5).
\end{eqnarray}
As a check of consistency, we consider the first equation in ~(\ref{Maxequ}), which leads to $ A_t'' (0) =   q^2 a_0+ \bar{\omega} q b_0$. It can be easily verified that our result~(\ref{At58}) is consistent.

Another check is from the Forbenius series of the equation of $A_t'$. The characteristic equation of~(\ref{at3p}) has two different integer roots, $u=0$ and $u=1$, which means that the general solution can be constructed by the following steps,
\begin{eqnarray}
A_t'& =& A Z_1 + B Z_2,\\
Z_1 &=& 1 + c_2 u^2 + o(u^4),\\
Z_2 &=& u + d_2 u^3 + o(u^4),
\end{eqnarray}
where $A=A_t'(0)$, $B=A_t''(0)$, $c_2= \frac{1}{2}(q^2-\bar{\omega}^2) $ and $d_{2}=\frac{1}{6}(q^2-\bar{\omega}^2)$.
Combing all the above results we arrive at
$A=a_1$, $B= q^2 a_0 + q \bar{\omega} b_0$.
According to~\cite{Son:2002sd}, the tt-component of the retarded Green's function is given by
\begin{equation}
G^{R}_{tt} = -M_{tt} ( 0) = - a_1 / a_0.
\end{equation}

Now let us consider the equation for $A_y$,
$$A_y'' + ( \frac{f'}{f} + \frac{b'}{b} )A_y' + \frac{1}{b^2 f^2} ( \bar{\omega}^2- q^2 b^2) A_y=0.
$$
It turns out that the asymptotic structure of the equation is similar to that of $A_t'$ up to the sixth order. Therefore we have the following solution for $A_{y}$
\begin{eqnarray}
A_y& =& \tilde{A}\tilde{Z}_1 + \tilde{B}\tilde{Z}_2,\\
\tilde{Z}_1 &=& 1 + \tilde{c}_2 u^2 + o(u^4),\\
\tilde{Z}_2 &=& u + \tilde{d}_2 u^3 + o(u^4),
\end{eqnarray}
where $\tilde{A}=A_y(0)$, $\tilde{B}=A_y'(0)$, $\tilde{c}_2= \frac{1}{2}(q^2-\bar{\omega}^2) $ and $\tilde{d}_{2}=\frac{1}{6}(q^2-\bar{\omega}^2)$.
Finally if we set $A_y(0) = a_1$ and $A_y'(0) = q^2 a_0$, we can obtain
\begin{equation}
G^{R}_{tt} G^{R}_{yy} = -  \frac{a_{1}}{a_{0}} ( q^2 \frac{a_0 }{a_1} ) = -q^2,
\end{equation}
which leads to
\begin{equation}
K^{L}(\omega,k)K^{T}(\omega,k)=1,
\end{equation}
by combining~(\ref{GRK}). This agrees with the result obtained in~\cite{Herzog:2007ij}, where the analysis was performed in the large $N$ limit. Therefore our result suggests that such a relation does not receive finite $N$ corrections.
\subsection{Adding Weyl corrections}
In this subsection we consider the case with Weyl corrections, where the EM self-duality is explicitly broken.
We start with the EM dual theory following~\cite{Myers:2010pk}
\begin{equation}
S_{\tilde{A}}=\int d^{4}x\sqrt{-g}(-\frac{1}{8\hat{g}_{4}^{2}}G_{\mu\nu}\hat{X}^{\mu\nu\rho\sigma}G_{\rho\sigma}),
\end{equation}
where $G_{\mu\nu}=\partial_{\mu}B_{\nu}-\partial_{\nu}B_{\mu}$ denotes the dual field strength and the gauge coupling in the dual theory is given by $\hat{g}_{4}^{2}=1/g_{4}^{2}$. The quantity $\hat{X}$ is given by
\begin{equation}
{\hat{X}_{\mu\nu}}^{\rho\sigma}=-\frac{1}{4}{\epsilon_{\mu\nu}}^{\lambda\eta}{(X^{-1})_{\lambda\eta}}^{\alpha\beta}
{\epsilon_{\alpha\beta}}^{\rho\sigma}.
\end{equation}
Note that ${X_{\mu\nu}}^{\rho\sigma}$ satisfies the following relation
\begin{equation}
\frac{1}{2}{(X^{-1})_{\mu\nu}}^{\lambda\eta}{X_{\lambda\eta}}^{\rho\sigma}\equiv{I_{\mu\nu}}^{\rho\sigma},
\end{equation}
where $\epsilon_{\mu\nu\rho\sigma}$ is the totally antisymmetric tensor. $G_{\mu\nu}$ is related to $F_{\mu\nu}$ by
\begin{equation}
\label{FGrelation}
F_{\mu\nu}=\frac{g_{4}^{2}}{4}{(X^{-1})_{\mu\nu}}^{\rho\sigma}{\epsilon_{\rho\sigma}}^{\lambda\eta}G_{\lambda\eta}.
\end{equation}
Moreover, in Maxwell theory we have $X^{-1}=X$, which yields ${\hat{X}_{\mu\nu}}^{\rho\sigma}={I_{\mu\nu}}^{\rho\sigma}$.
This is a demonstration that the Maxwell theory is self-dual.

When self-duality is lost, we generally find that $\hat{X}\neq X$. For notation convenience, let us define a six-dimensional space of antisymmetric index pairs
$$A,B\in\{tx, ty, tu, xy, xu, yu\},$$
which enables us to express $X$ in a simple form,
\begin{equation}
{X_{A}}^{B}={\rm diag}\left(X_{1}(u), X_{2}(u), X_{3}(u), X_{4}(u), X_{5}(u), X_{6}(u)\right).
\end{equation}
Note that due to rotational symmetry in the xy-plane, we have $X_{1}(u)=X_{2}(u), X_{5}(u)=X_{6}(u)$, however we will leave this symmetry implicit.
In addition, $X^{-1}$ is simply $1/X$ and the components of ${\hat{X}_{A}}^{B}$ are given by
\begin{eqnarray}
{\hat{X}_{A}}^{B}&=&{\rm diag}\left(\hat{X}_{1}(u), \hat{X}_{2}(u), \hat{X}_{3}(u), \hat{X}_{4}(u), \hat{X}_{5}(u), \hat{X}_{6}(u)\right)\nonumber\\
&=&{\rm diag}\left(\frac{1}{X_{6}}, \frac{1}{X_{5}},\frac{1}{X_{4}},\frac{1}{X_{3}},\frac{1}{X_{2}},\frac{1}{X_{1}}\right).
\end{eqnarray}
Using the notation $A, B$, the duality transformation of the field strengths~(\ref{FGrelation}) becomes
\begin{equation}
F_{A}=g_{4}^{2}{(X^{-1})_{A}}^{B}{\epsilon_{B}}^{C}G_{C}.
\end{equation}
Working in the background~(\ref{bhuzb}), the components of the field strength are given by
\begin{eqnarray}
\label{FGrelationcomponent}
& &F_{tx}=\frac{g_{4}^{2}}{X_{1}}r_{0}f b G_{yu},~~F_{ty}=-\frac{g_{4}^{2}}{X_{2}}r_{0}f b G_{xu},\nonumber\\
& &F_{tu}=\frac{g_{4}^{2}}{X_{3}}\frac{b}{r_{0}}G_{xy},
~~F_{xy}=-\frac{g_{4}^{2}}{X_{4}}\frac{r_{0}}{b }G_{tu},\nonumber\\
& &F_{xu}=\frac{g_{4}^{2}}{X_{5}}\frac{1}{r_{0}f b}G_{ty},~~
F_{yu}=-\frac{g_{4}^{2}}{X_{6}}\frac{1}{r_{0}f b}G_{tx}.
\end{eqnarray}

The equations of motion for $A_{\mu}$ become
\begin{eqnarray}
\label{eom6}
& &A_{t}^{\prime\prime}+(\frac{X_{3}^{\prime}}{X_{3}}-\frac{b^{\prime}}{b})A_{t}^{\prime}
-\frac{X_{1}}{X_{3}f}(\bar{\omega}qA_{x}+q^{2}A_{t})=0,\nonumber\\
& &A_{x}^{\prime\prime}+(\frac{X_{5}^{\prime}}{X_{5}}+\frac{f^{\prime}}{f}+\frac{b^{\prime}}{b})A_{x}^{\prime}
+\frac{X_{1}}{X_{5}}\frac{1}{b(u)^{2}f(u)^{2}}(\bar{\omega}^{2}A_{x}+\bar{\omega}qA_{t})=0,\nonumber\\
& &\bar{\omega}X_{3}A_{t}^{\prime}+qfb^{2}X_{5}A_{x}^{\prime}=0,\nonumber\\
& &A_{y}^{\prime\prime}+(\frac{X_{6}^{\prime}}{X_{6}}+\frac{f^{\prime}}{f}+\frac{b^{\prime}}{b})A_{y}^{\prime}
+\frac{1}{f^{2}X_{6}}(\frac{\bar{\omega}^{2}}{b ^{2}}X_{2}-q^{2}f X_{4})A_{y}=0,
\end{eqnarray}
where $\bar{\omega}=\omega/r_{0},~q=k/r_{0}$. For the EM dual theory, the equations of motion can be obtained by simply replacing
$A_{\mu}\rightarrow B_{\mu}, X_{i}\rightarrow\hat{X}_{i}$.

The retarded Green's function can be extracted following~\cite{Son:2002sd}. Expanding the action and integrating by parts leave a surface term at the asymptotic boundary,
\begin{equation}
S_{A}=\frac{r_{0}}{2g_{4}^{2}}\int d^{3}x\left(\frac{X_{3}}{b}A_{t}^{\prime}A_{t}-X_{5}f b A_{x}A_{x}^{\prime}
-x_{6}f b A_{y}A_{y}^{\prime}\right)\Big|_{u\rightarrow0}.
\end{equation}
Therefore the desired retarded Green's functions are given by
\begin{eqnarray}
& &G^{R}_{tt}=\frac{r_{0}}{g_{4}^{2}}\frac{X_{3}}{b}\frac{\delta A_{t}^{\prime}}{\delta A_{t}^{0}}\Big|_{u\rightarrow0},\nonumber\\
& &G^{R}_{yy}=-\frac{r_{0}}{g_{4}^{2}}X_{6}f b\frac{\delta A_{y}^{\prime}}{\delta A_{y}^{0}}\Big|_{u\rightarrow0}.
\end{eqnarray}
Next let us consider the Green's function $G^{R}_{yy}$ and assume $A_{y}(u)=\psi(u)A_{y}^{0}$ with the asymptotic normalization $\psi(0)=1$.
Then we can obtain
\begin{equation}
\label{GRyy}
G^{R}_{yy}=-\frac{r_{0}}{g_{4}^{2}}X_{6}(0)f(0)b(0)\psi^{\prime}(0),
\end{equation}
where we still keep $f(0)$ and $b(0)$ explicitly for generality.
Recall the relation between $F_{\mu\nu}$ and $G_{\mu\nu}$ in~(\ref{FGrelationcomponent}), in particular,
\begin{equation}
F_{xy}=-\frac{g_{4}^{2}}{X_{4}}\frac{r_{0}{b}}{G_{tu}}.
\end{equation}
Note that $G_{tu}=-B^{\prime}_{t},~~F_{xy}=ikA_{y}$, we have
\begin{equation}
B_{t}^{\prime}\propto b X_{4}A_{y},
\end{equation}
which means that $bX_{4}A_{y}$ provides a solution to the equation of motion for $B_{t}(u)$ in the EM dual theory.
In order to satisfy the desired boundary condition in the asymptotics, we introduce a new constant $C_{1}$ so that $B_{t}^{\prime}(u)=C_{1}X_{4} b \psi B_{t}^{0}$. The value of $C_{1}$ can be fixed by considering the equation of motion for
 $B_{t}$, which is in analogous to the corresponding one in~(\ref{eom6}),
 \begin{equation}
 C_{1}=\frac{k(\omega B_{x}^{0}+kB_{t}^{0})}{r_{0}^{2}f(0)b(0)X_{6}(0)\psi^{\prime}(0)},
 \end{equation}
 where we have used the relation $\hat{X}_{3}=1/X_{4},~\hat{X}_{1}=1/X_{6}$.

 Then the retarded Green's function of the EM dual version turns out to be
 \begin{eqnarray}
 \label{GHRtt}
 \hat{G}_{tt}^{R}&=&\frac{r_{0}}{\hat{g}_{4}^{2}}\frac{\hat{X}_{3}}{b}\frac{\delta B_{t}^{\prime}}
 {\delta B_{t}^{0}}\Big|_{u\rightarrow0}\nonumber\\
 &=&\frac{g_{4}^{2}}{r_{0}}\frac{k^{2}}{X_{6}(0)f(0)b(0)\psi^{\prime}(0)}.
 \end{eqnarray}
Therefore by combining~(\ref{GRyy}) and~(\ref{GHRtt}) and using the relation~(\ref{GRK}), we arrive at the following result
 \begin{equation}
 \hat{G}^{R}_{tt}G^{R}_{yy}=-k^{2} ~~\Rightarrow~~K^{T}(\omega, k)\hat{K}^{L}(\omega,k)=1.
 \end{equation}
 The EM dual version of the above relation can be obtained in a straightforward way. We can similarly consider an expression for $\hat{G}_{yy}^{R}$ and the counterpart of~(\ref{GHRtt}), then we have
 \begin{equation}
 \hat{K}^{T}(\omega,k)K^{L}(\omega,k)=1.
 \end{equation}
 It was shown that these relations hold in Schwarzschild-AdS background in~\cite{Myers:2010pk} and in Lifshitz background~\cite{Lemos:2011gy}, i.e. in the limit $N\rightarrow\infty$. Moreover, it was pointed out in~\cite{Myers:2010pk} that these expressions have the same form as that obtained from particle-vortex duality without self-duality in the condensed matter context. Here we show that such relations still hold at finite $N$.

The relation~(\ref{KLKHT}) has direct consequence on the conductivity, as illustrated in~\cite{Sachdev:2012prb, Sachdev:2013prb}. When taking the limit of vanishing momentum $k\rightarrow0$, we arrive at $K^{T}(\omega, k)=K^{L}(\omega, k)$, which is required by rotational invariance and also holds for the dual version. Therefore the conductivity and its dual satisfies
\begin{equation}
\sigma(\gamma,\omega)\hat{\sigma}(\gamma,\omega)=1,
\end{equation}
which can be obtained by combining~(\ref{GRK}) and $G^{R}_{yy}=-i\omega\sigma(\omega,k)$. Moreover, the dual tensor $\hat{X}$ has the following expansion at small $\gamma$,
\begin{eqnarray}
{\hat{X}_{\mu\nu}}^{\rho\sigma}&=&{I_{\mu\nu}}^{\rho\sigma}+8\gamma{C_{\mu\nu}}^{\rho\sigma}+\mathcal{O}(\gamma^{2})\nonumber\\
&=&{X_{\mu\nu}}^{\rho\sigma}|_{\gamma\rightarrow-\gamma}+\mathcal{O}(\gamma^{2}).
\end{eqnarray}
Hence
\begin{equation}
\sigma(\gamma,\omega)\sigma(-\gamma,\omega)\approx1,~~|\gamma|\ll1.
\end{equation}
The above analysis was performed in Schwarzschild-${\rm AdS}_{4}$ background in~\cite{Sachdev:2012prb, Sachdev:2013prb}. Here it can be easily shown that those relations still hold. So we may conclude that S-duality interchange the locations of the conductivity zeros and poles even in the presence of higher derivative corrections to gravity.

\section{Summary}
The AdS/CFT correspondence in the usual sense lies in the limit $N\rightarrow\infty$, where the dual gravity description is given by Einstein gravity. To study physics at finite $N$, we should consider higher derivative corrections to Einstein theory in the dual gravity side. One extensively studied example is the Gauss-Bonnet gravity, which leads to many interesting phenomena. For example, it has been observed in~\cite{Gregory:2009fj} the formation of scalar hair becomes more difficult in Gauss-Bonnet theory. However, in $(3+1)$ dimensional bulk spacetimes, which are of great interest in AdS/CMT, the Gauss-Bonnet terms turn out to be trivial. Therefore in order to study finite $N$ effects in AdS/CMT, it is desirable to consider non-trivial higher derivative terms in $(3+1)$ dimensions.

Higher derivative effects in $AdS_{4}$ have been extensively investigated in~\cite{Smolic:2013gz}, where perturbative black hole solutions within $C^{3},~f(R)$ and $R^{4}$ corrections are obtained. As argued in~\cite{Smolic:2013gz}, the black hole solution within $C^{3}$ corrections can be regarded as the analogue of Gauss-Bonnet black hole in $(3+1)$ dimensions, which provides a natural playground for studying finite $N$ effects in AdS/CMT. In this paper we focus on holographic charge transport in this background and our main results are summarized as follows:
\begin{itemize}
\item We calculate the DC conductivity in both the standard Maxwell theory and in the presence of Weyl corrections. The DC conductivity in Maxwell theory is  the same as that at $N\rightarrow\infty$, while the Weyl-corrected conductivity depends on both the $C^{3}$ coupling $\alpha$ and the Weyl correction coupling $\gamma$.
\item The DC conductivity is calculated via the membrane paradigm and the Kubo's formula, which exhibit precise agreement.
\item We study the AC conductivity numerically and find that in the standard Maxwell theory, the conductivity is still a frequency-independent constant. The AC conductivity in the Weyl-corrected theory exhibits the desired properties obtained in CFT. Moreover, in the Weyl-corrected theory we find that the DS peak can be flattened as positive $\gamma$ decreases and the DC conductivity decreases when $\gamma$ approaches the lower bound $-1/12$ or $\alpha$ is bigger. It may be zero at certain $\alpha$ with fixed negative $\gamma$. This behavior may be seen as an analog of the crossover from a `metal' to a `bad' metal.
\item As $N\rightarrow\infty$, in standard Maxwell theory we have the following relation~\cite{Herzog:2007ij}
      $$K^{L}(\omega,k)K^{T}(\omega,k)=1,$$
      where $K^{L},~K^{R}$ are related to the tt- and yy-components of the retarded Green's function via~(\ref{GRK}). When EM self-duality is lost, we have~\cite{Myers:2010pk} $$K^{L}(\omega,k)\hat{K}^{T}(\omega,k)=1,$$where $\hat{K}^{T}$ denotes the counterpart in the EM-dual theory. We show that the above two relations still hold at finite $N$.
\end{itemize}
Even if the background is at finite temperature and zero density, our results suggest that we may construct a holographic dual of the metal-insulator transition in higher derivative theory. The holographic metal-insulator transition was realized in~\cite{Donos:2012js} by considering a five-dimensional Einstein-Maxwell-Proca theory with a helical lattice which breaks translation invariance. Recently the metal-insulator transition in 4D Einstein-Maxwell theory coupled to a complex scalar was realized in~\cite{Donos:2013eha}, where the translation symmetry is broken by the complex scalar. It is expected that the metal-insulator transition may be realized by considering charged black hole backgrounds in the presence of higher derivative corrections of the type~(\ref{CF2}).

{\bf Note added} We became aware of~\cite{William:2013wey} at the final stage of this work, where the author considers a large class of higher-derivative terms involving 2 powers of the field strength such as $C^2F^2$, etc. Some conclusions similar to ours are obtained.

\bigskip \goodbreak \centerline{\bf Acknowledgments}
\noindent
We thank Subir Sachdev, Marika Taylor and  William Witczak-Krempa for reading the manuscript and giving valuable comments. SB is supported by the DFG-grant SFB/Transregio 7 ¡°Gravitational Wave Astronomy¡±. DWP is supported by Alexander von Humboldt Foundation.

\appendix
\section{Bounds on $\gamma$}
In the appendix we show that the permitted range of $\gamma$ is still $\gamma\in[-1/12, 1/12]$ by considering the causal constraints of the boundary CFT.
 In other words, the presence of $\alpha$ will not affect the permitted value of $\gamma$. Similar analysis was also carried out in~\cite{Ritz:2008kh, Brigante:2008gz, Buchel:2009tt} and we will follow the steps in~\cite{Myers:2010pk}.

 The equations of motion for the gauge field are given as follows:
\begin{eqnarray}
& &A_{t}^{\prime\prime}+\frac{h_{t}^{\prime}}{h_{t}}A_{t}^{\prime}-P_{t}(\omega kA_{x}+k^{2}A_{t})=0,\label{appeomat}\\
& &\omega A_{t}^{\prime}+k P_{tx}A_{x}^{\prime}=0, \label{appeomcons}\\
& &A_{x}^{\prime\prime}+\frac{h_{x}^{\prime}}{h_{x}}+\frac{1}{r_{0}^{2}b^{2}f^{2}}(\omega^{2}A_{x}+k^{2}A_{t})=0,\label{appeomax}\\
& &A_{y}^{\prime\prime}+\frac{h_{y}^{\prime}}{h_{y}}A^{\prime}_{y}+(\omega^{2}P_{y1}+k^{2}P_{y2})A_{y}=0,\label{appeo}
\end{eqnarray}
where
\begin{eqnarray}
h_{t}&=&-\frac{r_{0}}{b}\left(1+\frac{4\gamma u^{2}}{3b}(3b^{\prime}f^{\prime}+2fb^{\prime\prime}+bf^{\prime\prime})\right),\nonumber\\
h_{x}&=&r_{0}bf\left(1-\frac{2\gamma u^{2}}{3b}(3b^{\prime}f^{\prime}+2fb^{\prime\prime}+bf^{\prime\prime})\right),\nonumber\\
h_{y}&=&r_{0}bf\left(1-\frac{2\gamma u^{2}}{3b}(3b^{\prime}f^{\prime}+2fb^{\prime\prime}+bf^{\prime\prime})\right),\nonumber\\
P_{t}&=&\frac{1}{r_{0}^{2}f}\frac{1-\frac{2\gamma u^{2}}{3b}(3b^{\prime}f^{\prime}+2fb^{\prime\prime}+bf^{\prime\prime})}{1+\frac{4\gamma u^{2}}{3b}(3b^{\prime}f^{\prime}+2fb^{\prime\prime}+bf^{\prime\prime})},\nonumber\\
P_{tx}&=&\frac{b^{2}f[2\gamma u^{2}(3b^{\prime}f^{\prime}+2fb^{\prime}+b(3-2\gamma u^{2}f^{\prime\prime}))]}{4\gamma u^{2}(3b^{\prime}f^{\prime}+2fb^{\prime\prime})+b(3+4\gamma u^{2}f^{\prime\prime})},\nonumber\\
P_{y1}&=&\frac{1}{r_{0}^{2}b^{2}f^{2}},\nonumber\\
P_{y2}&=&\frac{4\gamma u^{2}\left(3b^{\prime}f^{\prime}+2fb^{\prime\prime}+b(3+4\gamma u^{2}f^{\prime\prime})\right)}{r_{0}^{2}f\left(2\gamma u^{2}(3b^{\prime}f^{\prime}+2fb^{\prime\prime})+b(-3+2\gamma u^{2}f^{\prime\prime})\right)}.\nonumber
\end{eqnarray}
Using~(\ref{appeomat}) and~(\ref{appeomcons}) we can obtain a third order equation for $A_{t}$ in terms of $u$,
\begin{eqnarray}
& &A_{t}^{\prime\prime\prime}+g_{1}A_{t}^{\prime\prime}+g_{2}A_{t}^{\prime}=0,\label{appendAtppp}
\end{eqnarray}
where
\begin{eqnarray}
& &g_{1}=\frac{h_{t}^{\prime}}{h_{t}}-\frac{P_{t}^{\prime}}{P_{t}},\nonumber\\
& &g_{2}=\left(\frac{h_{t}^{\prime}}{h_{t}}\right)^{\prime}-\frac{P_{t}^{\prime}}{P_{t}}\frac{h_{t}^{\prime}}{h_{t}}-P_{t}\left(\frac{\omega k}{P_{tx}}+k^{2}\right).\nonumber
\end{eqnarray}
Consider the coordinate transformation
\begin{equation}
\label{dzdu}
\frac{dz}{du}=\frac{3}{fb},
\end{equation}
and the variable substitutions
\begin{eqnarray}
& &A_{t}^{\prime}(u)=G_{1}(u)\psi_{1}(u),\\
& &A_{y}(u)=G_{2}(u)\psi_{2}(u),
\end{eqnarray}
we can convert the equations of motion~(\ref{appendAtppp}) for $A_{t}^{\prime}$  and~(\ref{appeo}) for $A_{y}$ into Schr\"{o}dinger forms,
\begin{eqnarray}
& &-\partial_{z}^{2}\psi_{1}(z)+V(z)\psi_{1}(z)=\bar{\omega}^{2}\psi_{1}(z),\label{equpsi1}\\
& &-\partial_{z}^{2}\psi_{2}(z)+W(z)\psi_{2}(z)=\bar{\omega}^{2}\psi_{2}(z),\label{equpsi2}
\end{eqnarray}
where $\bar{\omega}=\omega/r_{0}$.
Here $G_{1}(u)$ and $G_{2}(u)$  are determined by certain lengthy first order differential equations.

The effective potentials $V$ and $W$  can be written in terms of $u$  as
\begin{eqnarray*}
V(u)=q^{2}V_{0}(u)+V_{1}(u),\\
W(u)=q^{2}W_{0}(u)+W_{1}(u),
\end{eqnarray*}
where $q=k/r_{0}$.
In the limit $q\rightarrow\infty$,  $V_{0}$ and $W_0$ are dominant terms in the respective potentials and we can solve for~(\ref{equpsi1}) and~(\ref{equpsi2}) in WKB approximations. To avoid the causal violation in the dual CFT, we should ensure that $V_{0}$ and $W_0$ are not larger than one in $u\in[0,1]$.
Firstly, expanding $V_{0}$ and $W_0$ near the boundary $u=0$, we have
\begin{eqnarray}
V_{0}(u)\simeq1+(-1-8\alpha+12\gamma+96\alpha\gamma)u^{3}+\cdots,\\
W_{0}(u)\simeq1+(-1-8\alpha-12\gamma-96\alpha\gamma)u^{3}+\cdots.
\end{eqnarray}
It means that if $-\frac{1}{12}\leq\gamma\leq\frac{1}{12}$,  the values of $V_0$ and $W_0$ are decreasing  away from $u=0$. It is easy to observe that $u=0$ is the location of the maximum values of $V_0$ and $W_0$. Indeed motivated by~\cite{William:2013wey}, since $V_0(u)  W_0(u) >0$ and $V_0(0) = W_0(0) =1$, $V_0$ and $W_0$ should be positive in $u\in[0,1]$.

In sum, in the presence of higher derivative corrections, the permitted range of $\gamma$ is given by
\begin{equation}
-\frac{1}{12}\leq\gamma\leq\frac{1}{12}.
\end{equation}

\begin{figure}
\begin{center}
\vspace{1cm}
\hspace{0.5cm}
\includegraphics[angle=0,width=0.45\textwidth]{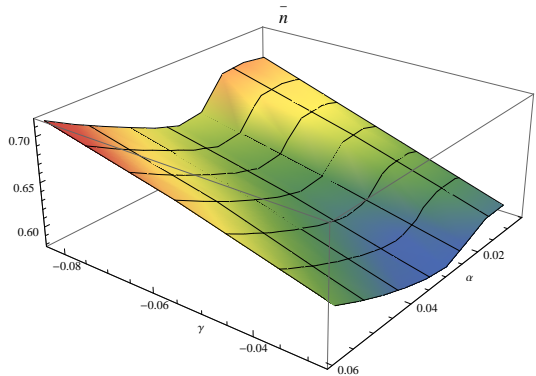}
\caption{The behavior of $\bar{n}$ in the range $-1/12\leq\gamma\leq0, 0\leq\alpha\leq0.06$. The maximal value of $\bar{n}$ is $0.7166$.}
\label{v1app}
\end{center}
\end{figure}

Next, we consider the instabilities. In the WKB regime, the fact that $V_0$ is non-negative indicates that it is not possible to have instabilities in the neutral plasma. Beyond the WKB regime, in particular for small momentum, we need to consider the contribution of $V_{1}$ to the potential $V$.
$V_1$ may achieve negative values close to the horizon for some negative $\gamma$ and then some unstable modes may appear in this regime. Following~\cite{Myers:2010pk},
a zero energy bound state can appear in  the potential well when
\begin{eqnarray}
(n-\frac{1}{2})\pi&\approx&\int^{\infty}_{z_{0}}dz\sqrt{-V_{1}(z)}\nonumber\\
&=&\int^{1}_{u_{0}}\frac{3du}{f(u)b(u)}\sqrt{-V_{1}(u)}\equiv I,
\end{eqnarray}
where $n$ is a positive integer and the integration is over the range of $u$ in which $V_{1}<0$. The result for $\bar{n}\equiv I/\pi+1/2$ is given in Fig.~\ref{v1app}. The maximal value of $\bar{n}=0.7166$ indicates that the potential well cannot support a negative energy bound state and there isn't any unstable mode appearing in the low momentum regime.

We have discussed about the large and small momenta limit of the potential $V$. Furthermore, for any finite momentum and for the parameter regime $\gamma \in [-1/12, 1/12]$ and $\alpha \in [0,0.06]$,  there are no instabilities from longitudinal vector mode because the positive contribution coming from $V_0$ compensates  the  negative contribution  from $V_1$.  On the other hand, to study the possible instabilities in the transverse vector mode, we perform a similar analysis on $W_0$ and observe that there are no additional instabilities in the parameter regime. Therefore although both the transverse and longitudinal modes of the vector exhibit instabilities, they only appear outside of the concerned parameter regime.

\newpage


\begin{thebibliography}{99}
\addcontentsline{toc}{section}{References}


\bibitem{Maldacena:1997re}
   J.~M.~Maldacena,
   ``The large N limit of superconformal field theories
   and supergravity'',
   Adv.\ Theor.\ Math.\ Phys.\  {\bf 2}, 231 (1998)
   [Int.\ J.\ Theor.\ Phys.\  {\bf 38}, 1113 (1999)]
   [arXiv:hep-th/9711200].
\bibitem{GKP:1998}
   S.~S.~Gubser, I.~R.~Klebanov and A.~M.~Polyakov,
   ``Gauge theory correlators from non-critical string theory'',
   Phys.\ Lett.\  B {\bf 428}, 105 (1998)
   [arXiv:hep-th/9802109].
\bibitem{Witten:1998}
   E.~Witten,
   ``Anti-de Sitter space and holography'',
   Adv.\ Theor.\ Math.\ Phys.\  {\bf 2}, 253 (1998)
   [arXiv:hep-th/9802150].

\bibitem{Aharony:1999ti}
   O.~Aharony, S.~S.~Gubser, J.~M.~Maldacena, H.~Ooguri and Y.~Oz,
   ``Large N field theories, string theory and gravity'',
   Phys.\ Rept.\  {\bf 323}, 183 (2000)
   [arXiv:hep-th/9905111].
\bibitem{Hartnoll:2009sz}
   S.~A.~Hartnoll,
   ``Lectures on holographic methods for condensed matter physics'',
   Class.\ Quant.\ Grav.\  {\bf 26}, 224002 (2009)
   [arXiv:0903.3246 [hep-th]].
\bibitem{Herzog:2009}
   C.~P.~Herzog,
   ``Lectures on Holographic Superfluidity and Superconductivity'',
   J.\ Phys.\ A  {\bf 42}, 343001 (2009)
   [arXiv:0904.1975 [hep-th]].
\bibitem{McGreevy:2009}
    J.~McGreevy,
   ``Holographic duality with a view toward many-body physics'',
   arXiv:0909.0518 [hep-th].
\bibitem{Horowitz:2010}
   G.~T.~Horowitz,
   ``Introduction to Holographic Superconductors'',
   arXiv:1002.1722 [hep-th].
\bibitem{Sachdev:2010}
   S.~Sachdev,
   ``Condensed matter and AdS/CFT'',
   arXiv:1002.2947 [hep-th].
\bibitem{Hartnoll:2011}
    S.~A.~Hartnoll,
  ``Horizons, holography and condensed matter,''
  arXiv:1106.4324 [hep-th].
\bibitem{Cai:2001dz}
  R.~-G.~Cai,
  ``Gauss-Bonnet black holes in AdS spaces,''
  Phys.\ Rev.\ D {\bf 65}, 084014 (2002)
  [hep-th/0109133].
\bibitem{Gregory:2009fj}
  R.~Gregory, S.~Kanno and J.~Soda,
  ``Holographic Superconductors with Higher Curvature Corrections,''
  JHEP {\bf 0910}, 010 (2009)
  [arXiv:0907.3203 [hep-th]].
\bibitem{Smolic:2013gz}
  J.~Smolic and M.~Taylor,
  ``Higher derivative effects for 4d AdS gravity,''
  JHEP {\bf 1306}, 096 (2013)
  [arXiv:1301.5205 [hep-th]].
\bibitem{Sachdev:2011qpt}
S.~Sachdev, ``Quantum Phase Transitions.'' Cambridge University Press, UK, 2nd edition, (2011).
\bibitem{Sachdev:1997prb}
S.~Sachdev,
``Nonzero-temperature transport near fractional quantum Hall critical points,''
Phys.\ Rev.\ D {\bf 57}, 7157 (1998) [cond-mat/9709243].
\bibitem{Damle:1997prb}
 K.~Damle and S.~Sachdev,
  ``Non-zero temperature transport near quantum critical point,''
  Phys.\ Rev.\ B {\bf 56}, 8714 (1997)
  [cond-mat/9705206].
\bibitem{Herzog:2007ij}
  C.~P.~Herzog, P.~Kovtun, S.~Sachdev and D.~T.~Son,
  ``Quantum critical transport, duality, and M-theory,''
  Phys.\ Rev.\ D {\bf 75}, 085020 (2007)
  [hep-th/0701036].
\bibitem{Hofman:2008ar}
  D.~M.~Hofman and J.~Maldacena,
  ``Conformal collider physics: Energy and charge correlations,''
  JHEP {\bf 0805}, 012 (2008)
  [arXiv:0803.1467 [hep-th]].
\bibitem{Hofman:2009ug}
  D.~M.~Hofman,
  ``Higher Derivative Gravity, Causality and Positivity of Energy in a UV complete QFT,''
  Nucl.\ Phys.\ B {\bf 823}, 174 (2009)
  [arXiv:0907.1625 [hep-th]].
\bibitem{Ritz:2008kh}
  A.~Ritz and J.~Ward,
  ``Weyl corrections to holographic conductivity,''
  Phys.\ Rev.\ D {\bf 79}, 066003 (2009)
  [arXiv:0811.4195 [hep-th]].
\bibitem{Myers:2010pk}
  R.~C.~Myers, S.~Sachdev and A.~Singh,
  ``Holographic Quantum Critical Transport without Self-Duality,''
  Phys.\ Rev.\ D {\bf 83}, 066017 (2011)
  [arXiv:1010.0443 [hep-th]].
\bibitem{Sachdev:2012prb}
 W.~Witczak-Krempa and S.~Sachdev,
  ``The quasi-normal modes of quantum criticality,''
  Phys.\ Rev.\ B {\bf 86}, 235115 (2012)
  [arXiv:1210.4166[cond-mat.str-el]].
\bibitem{Sachdev:2013prb}
 W.~Witczak-Krempa and S.~Sachdev,
  ``Dispersing quasinormal modes in (2+1)-dimensional conformal field theories,''
  Phys.\ Rev.\ B {\bf 87}, 155149 (2013)
  [arXiv:1302.0847[cond-mat.str-el]].
\bibitem{Iqbal:2008by}
  N.~Iqbal and H.~Liu,
  ``Universality of the hydrodynamic limit in AdS/CFT and the membrane paradigm,''
  Phys.\ Rev.\ D {\bf 79}, 025023 (2009)
  [arXiv:0809.3808 [hep-th]].
\bibitem{Son:2002sd}
  D.~T.~Son and A.~O.~Starinets,
  ``Minkowski space correlators in AdS / CFT correspondence: Recipe and applications,''
  JHEP {\bf 0209}, 042 (2002)
  [hep-th/0205051].
\bibitem{CaronHuot:2006te}
  S.~Caron-Huot, P.~Kovtun, G.~D.~Moore, A.~Starinets and L.~G.~Yaffe,
  ``Photon and dilepton production in supersymmetric Yang-Mills plasma,''
  JHEP {\bf 0612}, 015 (2006)
  [hep-th/0607237].
\bibitem{Atmaja:2008mt}
  A.~Nata Atmaja and K.~Schalm,
  ``Photon and Dilepton Production in Soft Wall AdS/QCD,''
  JHEP {\bf 1008}, 124 (2010)
  [arXiv:0802.1460 [hep-th]].

\bibitem{Imada:1998rmp}
M.~Imada, A.~Fujimori and Y.~Tokura, ``Metal-insulator transitions,'' Rev. Mod. Phys. {\bf 70}, 1039 (1998).
\bibitem{Dobro:2011oup}
V.~Dobrosavljevic, ``Introduction to metal-insulator transitions,'' in {\it Conductor Insulator Quantum Phase Transitions,} eds.
V.~Dobrosavljevic, N.~Trivedi and J.~M.~Valles Jr., OUP 2012 [arXiv:1112.6166 [cond-mat.str-el]].
\bibitem{Gun:2003rmp}
O.~Gunnarsson, M.~Calandra and J.~E.~Han, ``Colloquium: Saturation of electrical resistivity,'' Rev. Mod. Phys. {\bf 75}, 1085 (2003) [cond-mat/0305412].
\bibitem{Hus:2004phi}
N.~E.~Hussey, K.~Takenaka and H.~Takagi, ``Universality of the Mott-Ioffe-Regel limit in metals,'' Phil. Mag. {\bf 84}, 2847 (2004) [cond-mat/0404263].

\bibitem{Lemos:2011gy}
  J.~P.~S.~Lemos and D.~-W.~Pang,
  ``Holographic charge transport in Lifshitz black hole backgrounds,''
  JHEP {\bf 1106}, 122 (2011)
  [arXiv:1106.2291 [hep-th]].
\bibitem{Brigante:2008gz}
  M.~Brigante, H.~Liu, R.~C.~Myers, S.~Shenker and S.~Yaida,
  ``The Viscosity Bound and Causality Violation,''
  Phys.\ Rev.\ Lett.\  {\bf 100}, 191601 (2008)
  [arXiv:0802.3318 [hep-th]].
\bibitem{Buchel:2009tt}
  A.~Buchel and R.~C.~Myers,
  ``Causality of Holographic Hydrodynamics,''
  JHEP {\bf 0908}, 016 (2009)
  [arXiv:0906.2922 [hep-th]].
\bibitem{Donos:2012js}
  A.~Donos and S.~A.~Hartnoll,
  ``Interaction-driven localization in holography,''
  Nature Phys.\  {\bf 9}, 649 (2013)
  [arXiv:1212.2998 [hep-th]].
\bibitem{Donos:2013eha}
  A.~Donos and J.~P.~Gauntlett,
  ``Holographic Q-lattices,''
 JHEP {\bf 1404}, 040 (2014)
  [arXiv:1311.3292 [hep-th]].
\bibitem{William:2013wey}
 W.~Witczak-Krempa, ``Quantum critical charge response from higher derivatives: is more different?'' 
  Phys.\ Rev.\ B {\bf 89}, 161114(R), arXiv:1312.3334 [cond-mat.str-el].
\end{thebibliography}
\end{document}